\documentclass[10pt,conference]{IEEEtran}
\usepackage{cite}
\usepackage{amsmath,amssymb,amsfonts}
\usepackage{algorithmic}
\usepackage{graphicx}
\usepackage{textcomp}
\usepackage[dvipsnames, svgnames, x11names]{xcolor}
\usepackage[hyphens]{url}
\usepackage{fancyhdr}
\usepackage{hyperref}
\usepackage{enumitem}
\usepackage{array}
\usepackage{multirow}
\usepackage{makecell}
\usepackage{listings}
\usepackage{textcomp}
\newcommand{\explain}[1]{\textcolor{black}{#1}}
\newcommand{\future}[1]{\textcolor{black}{#1}}

\newcommand{\lp}[1]{\textcolor{black}{#1}}

\newcommand{\sgarch}[1]{\textcolor{black}{#1}}
\newcommand{\grs}[1]{\textcolor{black}{#1}}
\newcommand{\coreg}[1]{\textcolor{black}{#1}}
\newcommand{\proof}[1]{\textcolor{black}{#1}}
\newcommand{\SCMS}[1]{\textcolor{black}{#1}}
\newcommand{\chipletg}[1]{\textcolor{black}{#1}}
\newcommand{\hpcayear}{2024}

\newcommand{\hpcasubmissionnumber}{119}
\title{Gemini: Mapping and Architecture Co-exploration for Large-scale DNN Chiplet Accelerators}

\def\hpcacameraready{} 

\newcommand\hpcaauthors{Jingwei Cai$^{\dagger}$, Zuotong Wu$^{\ddagger\S}$, Sen Peng$^{\ddagger\S}$, Yuchen Wei$^{\dagger}$, Zhanhong Tan$^{\dagger}$, Guiming Shi$^{\dagger}$,\\ Mingyu Gao$^{\dagger\P\textasteriskcentered}$ and Kaisheng Ma$^{\dagger\textasteriskcentered}$}
\newcommand\hpcaaffiliation{Tsinghua University$^{\dagger}$, Xi'an Jiaotong University$^{\ddagger}$, IIISCT$^{\S}$, Shanghai AI Laboratory$^{\P}$ \\Corresponding Author$^{\textasteriskcentered}$}
\newcommand\hpcaemail{\{caijw21,gaomy,kaisheng\}@tsinghua.edu.cn}

\def\aeopen{}           
\def\aereviewed{}     
\def\aereproduced{} 

\author{
  \ifdefined\hpcacameraready
    \IEEEauthorblockN{\hpcaauthors{}}
      \IEEEauthorblockA{
        \hpcaaffiliation{} \\
        \hpcaemail{}
      }
  \else
    \IEEEauthorblockN{\normalsize{HPCA \hpcayear{} Submission
      \textbf{\#\hpcasubmissionnumber{}}} \\
      \IEEEauthorblockA{
        Confidential Draft \\
        Do NOT Distribute!!
      }
    }
  \fi 
}

\fancypagestyle{camerareadyfirstpage}{%
  \fancyhead{}
  
  \fancyhead[C]{
    \ifdefined\aeopen
    \parbox[][12mm][t]{13.5cm}{\hpcayear{} IEEE International Symposium on High-Performance Computer Architecture (HPCA)}    
    \else
      \ifdefined\aereviewed
      \parbox[][12mm][t]{13.5cm}{\hpcayear{} IEEE International Symposium on High-Performance Computer Architecture (HPCA)}
      \else
      \ifdefined\aereproduced
      \parbox[][12mm][t]{13.5cm}{\hpcayear{} IEEE International Symposium on High-Performance Computer Architecture (HPCA)}
      \else
      \parbox[][0mm][t]{13.5cm}{\hpcayear{} IEEE International Symposium on High-Performance Computer Architecture (HPCA)}
    \fi 
    \fi 
    \fi 
    \ifdefined\aeopen 
      \includegraphics[width=12mm,height=12mm]{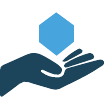}
    \fi 
    \ifdefined\aereviewed
      \includegraphics[width=12mm,height=12mm]{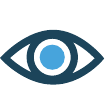}
    \fi 
    \ifdefined\aereproduced
      \includegraphics[width=12mm,height=12mm]{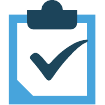}
    \fi
  }
  \fancyfoot[C]{}
}
\begin{document}
\maketitle

\ifdefined\hpcacameraready 
  \thispagestyle{camerareadyfirstpage}
  \pagestyle{empty}
\else
  \thispagestyle{plain}
  \pagestyle{plain}
\fi

\newcommand{\hpcaheight}{0mm}
\ifdefined\eaopen
\renewcommand{\hpcaheight}{12mm}
\fi


\begin{abstract}
Chiplet technology enables the integration of an increasing number of transistors on a single accelerator with higher yield in the post-Moore era, addressing the immense computational demands arising from rapid AI advancements. However, it also introduces more expensive packaging costs and costly Die-to-Die (D2D) interfaces, which require more area, consume higher power, and offer lower bandwidth than on-chip interconnects. Maximizing the benefits and minimizing the drawbacks of chiplet technology is crucial for developing large-scale DNN chiplet accelerators, which poses challenges to both architecture and mapping. Despite its importance in the post-Moore era, methods to address these challenges remain scarce.

To bridge the gap, we first propose a layer-centric encoding method to encode Layer-Pipeline (LP) spatial mapping for large-scale DNN inference accelerators and depict the optimization space of it. Based on it, we analyze the unexplored optimization opportunities within this space, which play a more crucial role in chiplet scenarios. Based on the encoding method and a highly configurable and universal hardware template, we propose an architecture and mapping co-exploration framework, Gemini, to explore the design and mapping space of large-scale DNN chiplet accelerators while taking monetary cost (MC), performance, and energy efficiency into account. Compared to the state-of-the-art (SOTA) Simba architecture with SOTA Tangram LP Mapping, Gemini's co-optimized architecture and mapping achieve, on average, 1.98$\times$ performance improvement and 1.41$\times$ energy efficiency improvement simultaneously across various DNNs and batch sizes, with only a 14.3\% increase in monetary cost. Moreover, we leverage Gemini to uncover intriguing insights into the methods for utilizing chiplet technology in architecture design and mapping DNN workloads under chiplet scenarios. \textit{The Gemini framework is open-sourced at \url{https://github.com/SET-Scheduling-Project/GEMINI-HPCA2024}}.

\end{abstract}

 \vspace{-1mm}
\section{Introduction}

\label{ch1:intro}
\noindent As Deep Neural Networks (DNNs) tackle increasingly complex problems, their size and complexity grow rapidly, resulting in increased computing and storage demands~\cite{DPN,bert,transformer,inception-v4,resnetxt}. While applying more advanced technology and enlarging single chip sizes have led to the development of many large-scale monolithic accelerators with tens of billions of transistors~\cite{graphcore,tenstorrent,TPUlesson,hanguang}, the end of Moore's Law~\cite{moorelaw} and limited photomask size pose significant challenges to further transistor integration.

Chiplet technology, using advanced packaging to combine small functional dies, offers a promising solution to overcome these limitations and enable continuous transistor integration. Chiplet-based DNN inference accelerators, such as Simba with 36 dies~\cite{simba}, have emerged. However, this technology introduces new challenges for architectural design and DNN mapping for large-scale chiplet accelerators, which are outlined below:


\textit{For architecture design}, the main challenge is determining the optimal chiplet granularity. While chiplet technology improves area limits and yield, it introduces higher packaging expenses and D2D interconnection costs. D2D interconnections are more energy and area-intensive, but provide lower bandwidth than on-chip lines. All the above adverse effects are collectively referred to as \textit{Chiplet Costs}. Consequently, A trade-off arises between using more smaller chiplets for better yield and fewer larger chiplets for lower \textit{Chiplet Costs}. Balancing this trade-off remains an unsolved challenge.

\textit{For DNN mapping}, the main challenges stem from the larger scale enabled by chiplet technology, and the costly D2D links. For the first challenge, maintaining high utilization and energy efficiency becomes increasingly difficult with the growing scale of accelerators. LP mapping, in which multiple layers are spatially mapped onto the accelerator, is widely employed by large-scale accelerators in both academia~\cite{tangram,scaledeep,isaac,simba} and industry~\cite{dojo,cerebras2,tenstorrent} to relieve the challenge. The core of LP mapping is spatial mapping (SPM), which determines which part of which layer is allocated to which core, and significantly impacts the performance and energy efficiency of large-scale accelerators. 
Despite the importance of LP SPM, most current strategies are still heuristic~\cite{tangram,scaledeep,tenstorrent,cerebras2,simba}. The problems and optimization space of LP SPM have not been clearly defined, exhaustively explored, or fully understood. This limitation constrains the ability to fully leverage optimization opportunities in LP mapping, which becomes increasingly important as the scale of accelerators and the structural complexity of DNNs increase. The second challenge lies in that D2D links tend to consume more energy and provide lower bandwidth than on-chip lines. Therefore, devising spatial mapping strategies that can automatically reduce D2D communications is vital for enhancing chiplet accelerator performance and efficiency, and fully harnessing the benefits provided by chiplet technology. However, there is a noticeable absence of such approaches in the existing literature.



The above analysis reveals that the employment of chiplet technology introduces intricate trade-offs and challenges. Thus, maximizing the benefits of chiplet technology and minimizing its disadvantages is crucial for developing DNN chiplet accelerators. This goal not only poses challenges for architectural design but also necessitates more efficient mapping schemes that effectively utilize larger accelerators while reducing the expensive communication costs between chiplets. To address these challenges, we have made the following contributions:

(1) We propose a layer-centric encoding method for representing LP SPM schemes in many-core chiplet DNN inference accelerators. Leveraging this encoding method, we delineate the optimization space for LP mapping, calculate its immense size, which significantly outstrips existing heuristic strategies, and analyze the latent optimization opportunities concealed within this space. These opportunities become particularly crucial under post-Moore chiplet times. To the best of our knowledge, this is \textbf{the first} work that clearly and systematically defines the optimization space of LP SPM for DNN inference accelerators.

(2) Based on the aforementioned encoding and a highly configurable and universal hardware template, we develop Gemini, a mapping and architecture co-exploration framework for large-scale DNN chiplet accelerators. Gemini includes two main engines: the Mapping Engine and the Monetary Cost Evaluator. In the Mapping Engine, a Simulated Annealing (SA) algorithm with five specifically-designed operators is developed to explore the extensive space defined by our encoding method and automatically minimize costly D2D communication. The Monetary Cost Evaluator assesses the MC of accelerators with varying architectural parameters. To the best of our knowledge, Gemini is \textbf{the first} framework to jointly explore the optimization space of mapping and architecture for large-scale DNN chiplet accelerators, considering not only \textcolor{black}{energy consumption} and performance but also MC. \textbf{The Gemini framework is open-sourced at \url{https://github.com/SET-Scheduling-Project/GEMINI-HPCA2024}.}

(3) We utilize Gemini to reveal several intriguing insights about utilizing chiplet technology in architecture design and mapping DNN workloads under chiplet scenarios as follows. (1) With the support of advanced mapping, moderately partitioning an accelerator into chiplets can reduce MC with nearly no loss of performance and energy efficiency. However, overly fine-grained chiplet partitions will simultaneously worsen MC, performance, and energy efficiency. (2) Regarding the granularity of computing cores, both energy efficiency and performance initially increase (albeit at a decelerating pace) as the granularity becomes finer (i.e., more cores). However, they experience a slight decline thereafter. Moreover, the MC tends to rise as the granularity of computing cores becomes finer, corresponding to an increase in the number of cores. (3) With the right design considerations and aided by the Gemini framework, employing a single chiplet for multiple accelerators becomes an effective approach for DNN inference accelerators. However, we also demonstrate the limitations of a ``one-size-fits-all'' method, particularly highlighting the impracticality of designing a small-scale chiplet intended to cater to a very wide spectrum of computational power requirements across diverse scenarios. (4) We study the property of spatial mapping, and find that rather than allocating chiplets for each layer in a gathered manner, gathering clusters with heavy data transfer together is more important and beneficial.

(4) Compared to the SOTA Simba architecture with Tangram SPM, Gemini's co-optimized architecture and mapping, on average, achieve 1.98$\times$ performance and 1.41$\times$ energy efficiency simultaneously across various DNNs and batch sizes, with only 14.3\% increase in MC.

 \begin{figure}[!b]

    \centering
    \includegraphics[width=3.3in]{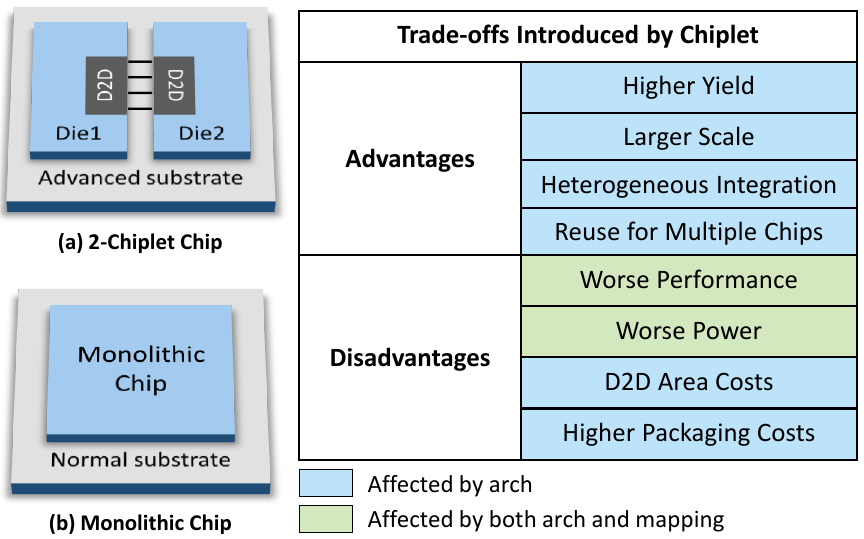}
    \caption{Trade-offs Introduced by Chiplet}
    \label{figure:tradeoff}

\end{figure}

\section{Background and Motivation}

\subsection{Trade-offs Introduced by Chiplet}\label{ch2_1:chiplet}

\noindent Chiplet technology is fundamentally an advanced packaging solution. It integrates multiple dies (chiplets) onto a substrate, which could be an organic substrate~\cite{zen4,zen3} or a silicon interposer~\cite{interposer_2500,interposer1}. This integration is achieved through high-density interconnections, allowing the dies to function collectively as a single chip. This technology offers numerous benefits, but it also incurs additional costs. In this section, we discuss the trade-offs of employing chiplet technology.

As shown in Fig.~\ref{figure:tradeoff}, the advantages of introducing chiplets are fourfold. Firstly, it increases the overall yield of a chip. By partitioning a large-scale monolithic chip into smaller chiplets, the overall yield of the chip can be significantly improved. For example, at the 7nm technology node, the yield of an 800 ${mm}^2$ chip and a 200 ${mm}^2$ chip are approximately 18\% and 75\%, respectively~\cite{chiplet_yinxiao}. Second, it extends the area limit of a chip, as advanced substrates (e.g., organic substrates~\cite{simba-jssc} or interposers~\cite{interposer_2500}) have a larger area limit than monolithic chips (858${mm}^2$~\cite{interposer_2500} of a reticle). Third, it enables heterogeneous integration. Unlike logic circuits, analog circuit IPs do not significantly benefit from the performance and density improvements brought about by technological advancements~\cite{zen2,zen3,pontevecchio,baton}. Therefore, manufacturing logic circuits with advanced technologies while producing various IO-functional analog IPs with more outdated processes can save on the expensive manufacturing and design and IP-related costs associated with advanced technologies~\cite{zen2,zen3,zen4,pontevecchio}. Fourth, the ability to repurpose a single chiplet for developing multiple computational chips, each differing in scale or targeted application, is a significant advantage of chiplet technology. This approach can substantially reduce the enormous Non-Recurring Engineering (NRE) costs and the time traditionally associated with developing different chips for each scale and scenario. A notable success story in this regard is AMD's Zen series CPUs~\cite{zen3,zen4}. While the advantages of chiplet technology are not the main focus of our research, given our primary emphasis on the architecture of an accelerator designed for a specific scenario, we do include a brief case study on this topic in Sec.~\ref{SCMS} for comprehensive understanding.


Fig.\ref{figure:tradeoff} also depicts the fourfold disadvantages of chiplets, all resulting from D2D interfaces. First, they increase \textcolor{black}{energy consumption}, as D2D links consume several to dozens of times more energy than the less than 0.1 pJ/bit data transfer cost with on-chip lines\cite{grs2019jssc,serdes2,serdes3,LVDS}. Second, limited inter-chiplet communication bandwidth may decrease performance. Compared to sufficient on-chip interconnect resources, interconnects between chiplets have limited bandwidth due to the limited number of IO pins available around each chiplet. Third, D2D interfaces require more area, as they need a specific analog PHY and controller, unlike on-chip lines, which occupy almost no silicon area. Fourth, the need for massive interconnections between chiplets increases packaging substrate costs, as advanced packaging scenarios require organic substrates with dozens of layers or silicon interposers, compared to the basic fan-out substrates sufficient for monolithic chips.


All the trade-offs above influence three facets of a chip: \textcolor{black}{energy consumption}, performance, and \textbf{MC}. It's particularly important to emphasize that in the chiplet era, solely considering the silicon chip area is inadequate for evaluating a chip's MC. Factors such as yield and packaging costs also need to be taken into account. Consequently, designing an efficient DNN chiplet accelerator requires carefully balancing among these trade-offs. However, existing works fall short in considering all these trade-offs simultaneously. Some works~\cite{chiplet_yinxiao,chiplet_cost} focus on studying the MC of chiplet-based chips, but do not consider their architectural details and the corresponding influence on performance and \textcolor{black}{energy consumption}. NN-baton~\cite{baton} focuses on optimizing \textcolor{black}{energy consumption} and performance for small-scale DNN chiplet accelerators but does not consider the trade-offs on MC. Furthermore, NN-baton's ring-based template and layer-sequential mapping strategies limit its applicability to small-scale accelerators and restrict its scalability.

The significant potential and intricate trade-offs of applying chiplet technology to DNN inference accelerators, alongside the current research void, inspire us to create a framework that co-explores architecture and mapping for these accelerators.


\subsection{Mapping Challenges}\label{ch2_2:mapping_challenge}

\noindent While chiplet technology facilitates the construction of larger-scale accelerators, translating theoretical computing power into actual computing performance presents an escalating challenge. A significant portion of existing research is focused on optimizing layer-sequential (LS) mapping for small-scale accelerators~\cite{timeloop,magnet,understand,tenet,hasco,gammatushka,mindmapping,CoSA,maeri,tetris}.However, with the increase in accelerator scale, LS mapping demonstrates limited scalability~\cite{tangram,simba,fused-layer,atomic}, whereas LP mapping exhibits a higher potential. For instance, Simba employs a naive LP mapping for some blocks of layers in ResNet-50, achieving approximately a $1.7\times$ throughput improvement over its original LS approach.  

\begin{figure*}[!thb]

     \centering
     \includegraphics[width=7in]{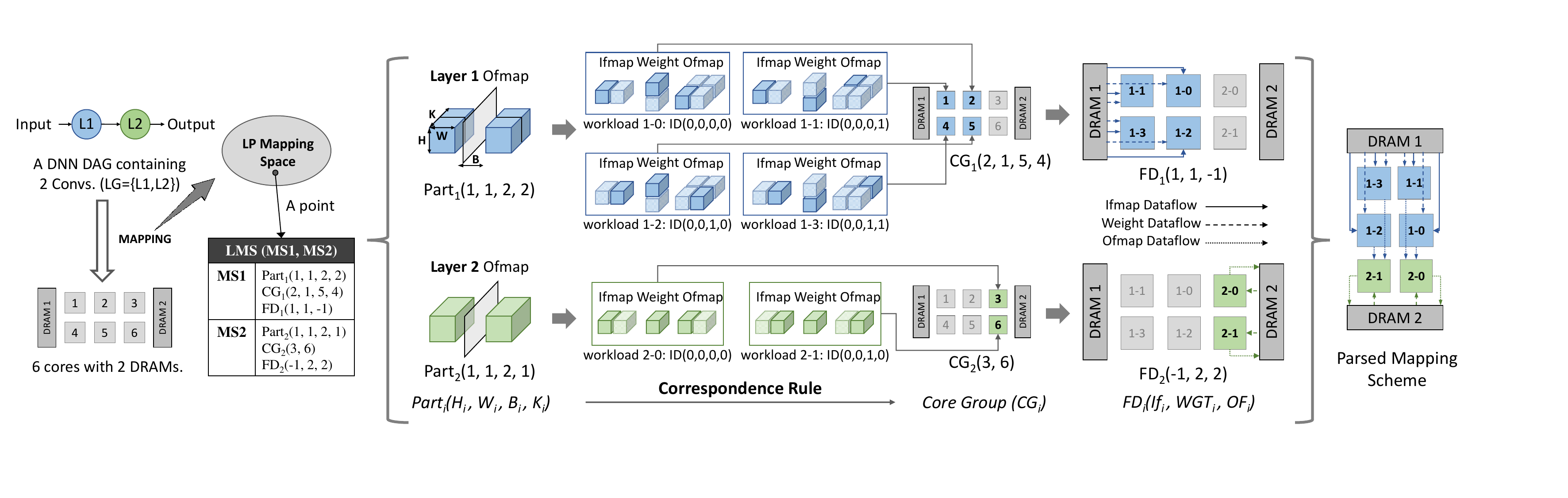}
     \caption{Parsing an Encoded LMS in the LP SPM Optimization Space into an Actual SPM Scheme}
     \label{figure:example}

 \end{figure*}

\lp{DNNs can be viewed as a Directed Acyclic Graph (DAG), where each layer is a node. In LP mapping, a graph (or subgraph) can be mapped onto a core array simultaneously, where different core groups compute different layers. The on-chip interconnect is responsible for transmitting the feature maps of layers that share dependencies. SPM determines which core specifically computes which part of which layers. As shown in Fig.~\ref{figure:example}, the top-left corner shows a DNN DAG containing two layers. In LS mapping, all six cores are used to compute each layer one by one, whereas in LP mapping, some of the six cores are used to compute the first layer, and the remaining cores are used for the second layer. The feature maps between the two layers can be transferred via the on-chip network without the need to access DRAM.}


While LP mapping presents significant potential, it is less extensively researched than LS mapping. Most existing studies employ techniques such as fine-grained pipelining~\cite{fused-layer,tangram} and temporal merging of some layers or workloads~\cite{SET,atomic} to reduce the imbalance across different pipeline stages and mitigate filling and draining overheads in LP. However, with regard to the SPM part, these studies do not offer specific optimizations and directly adopt heuristic stripe-based SPM strategies, which assign each layer to a consecutive and rectangle-shaped group of cores. The problem and its corresponding optimization space for LP SPM have not been clearly defined, let alone thoroughly explored or understood. It is worth noting that as most existing works also have an SPM optimization stage, the optimization method proposed in this work can easily integrate with existing methodologies.


Another significant challenge posed to SPM is mitigating the adverse effects on \textcolor{black}{energy consumption} and performance introduced by chiplet technology (Fig.~\ref{figure:tradeoff}). Fundamentally, these negative influences stem from the integration of high-energy-cost and lower-bandwidth D2D links into the chip. Consequently, this challenge primarily revolves around reducing D2D communication automatically for various workloads, an aspect that is not considered and optimized by existing heuristic LP SPM strategies~\cite{tangram,atomic,SET}.

\begin{figure}[!b]
\vspace{-1mm}
    \centering
    \includegraphics[width=3.3in]{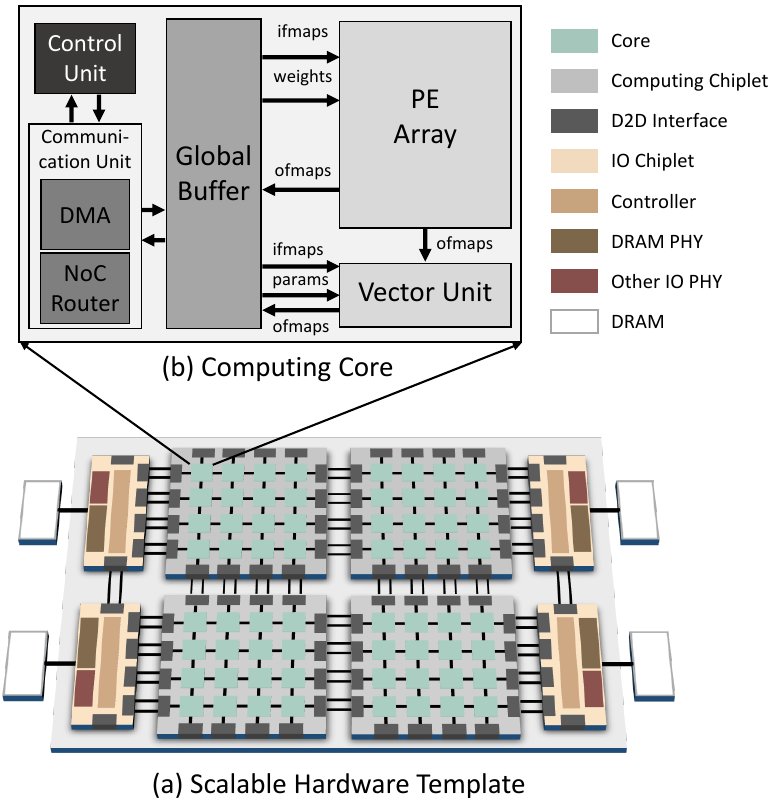}
    \caption{ Architecture of Scalable Hardware Template}
    \label{figure:hardware}

\end{figure}

\section{Scalable Hardware Template}\label{ch3:hardware_model}

\noindent In this section, we introduce the architecture of our universal and configurable hardware template, which is created by extracting common features from existing chiplet accelerators~\cite{simba,dojo} and large-scale accelerators~\cite{tenstorrent,NNP-I,tangram,tetris,cerebras2}.


\noindent\textbf{Overall Architecture}: As shown in Fig.~\ref{figure:hardware}(a), the proposed template comprises two distinct types of chiplets: IO chiplets and Computing chiplets. A mesh NoC interconnects all computing cores in all Computing chiplets and the controllers within IO chiplets, allowing for arbitrary core-to-core, core-to-DRAM, and DRAM-to-core communication. For inter-chiplet communication, the D2D transmission (TX) within a chiplet independently encodes the data and forwards it to the corresponding D2D reception (RX) in another chiplet. The D2D RX decodes the data and proceeds with NoC transmission. This inter-chiplet communication is fully automatic and transparent to both the source and the destination. 

This heterogeneous architecture allows for arbitrary numbers of IO and computing chiplets, ensuring the scalability of the entire template. Otherwise, if we equip each computing chiplet with IO-related PHY and controller would occupy the chip's edge area and IO pins, which could affect the routing of D2D links and, consequently, the system's scalability.

Our hardware template and the corresponding hardware model in Gemini Framework (see Sec.~\ref{ch5:gemini}) are not limited to supporting only mesh topology. They can be easily modified to support various topologies (demonstrated in Sec.~\ref{ch6_2:overall_compare}). However, in order to ensure the signal integrity of the D2D links, there are limitations on the interconnect distance and cross-linking, which restricts certain topologies like torus and butterfly. Therefore, it is preferable to maintain a point-to-point parallel interconnect, as shown in Fig.~\ref{figure:hardware}. Considering these factors and the fact that most existing tiled accelerators~\cite{dojo,simba,NNP-T,cerebras2} currently adopt mesh interconnect networks, we default to using the mesh topology in this paper. 



\noindent\textbf{Computing Chiplet Architecture}: Each Computing Chiplet contains an arbitrary number of computing cores interconnected by a mesh NoC. To enhance the scalability of this template, D2D interfaces are placed around the Chiplet, whose number is equal to the number of computing cores on each side. This arrangement enables the Computing Chiplet to form a larger-scale mesh with other Chiplets. In the example illustrated in Fig.~\ref{figure:hardware}(a), there are four cores on each side, so we place four D2Ds on each side of the Computing Chiplet.

The computing cores are the key components responsible for performing calculations in the entire accelerator, and their architecture is shown in Fig.~\ref{figure:hardware}(b). The communication unit comprises DMA and router, facilitating communication with other cores and DRAM. The control unit is primarily responsible for managing computation tasks based on statically-compiled instructions and task progress information, as well as managing the reception and transmission of data or messages to and from external cores or DRAMs. The global buffer (GLB) of each core is globally visible in the entire accelerator. Every core can read data from or write data to the GLBs of other cores, provided the data is valid, or the address can be written to. The PE array and the vector unit are responsible for computing the General Matrix Multiply (GEMM)/Convolution (Conv) and vector/scalar operators, respectively. Each time, the PE array reads a workload tile's weight and input feature maps (ifmaps) from the GLB. The resulting ofmaps/partial sum (psum) can be directly written back to the GLB or post-processed within the vector unit (such as Batch Normalization and ReLU operations). Simultaneously, the vector unit can be independently invoked to compute vector/scalar operators. 


\noindent\textbf{IO Chiplet Architecture}: The IO Chiplet is equipped with an array of IO functionalities, enabling interactions with DRAM, host systems, or other input sources (e.g., cameras). All input data from host systems or alternative input sources are first loaded into DRAM, and then can be loaded and processed by computing cores. The DRAM controller is also connected to multiple routers within the entire mesh NoC to match the bandwidth of the DRAM and the network, ensuring the full utilization of DRAM bandwidth.


\noindent\textbf{Configurable Parameters}: Our template features excellent configurability, offering a range of adjustable architectural parameters. These parameters include NoC bandwidth, D2D bandwidth, total DRAM bandwidth, the total number of cores in the X-direction (e.g., 8 in Fig.~\ref{figure:hardware}), the total number of cores in the Y-direction (e.g., 8 in Fig.~\ref{figure:hardware}), the number of chiplet divisions in the X-direction ($X_{Cut}$)(e.g., 2 in Fig.~\ref{figure:hardware}), the number of chiplet divisions in the Y-direction ($Y_{Cut}$)(e.g., 2 in Fig.~\ref{figure:hardware}), the number of MACs in the PE array within a single core, and the size of the GLB per core. It is worth mentioning that the microarchitecture of the PE array and its corresponding dataflow have been extensively studied in existing works~\cite{understand,eyeriss,magnet,interstellar,confuciuxtushka,mindmapping,baton}. Therefore, in this work, the PE array adopts the classic NVDLA architecture~\cite{nvdla,magnet} and corresponding dataflow to maintain a fair comparison with the baseline, Simba. \textbf{\textit{Of course, NVDLA can also be replaced by other microarchitectures with different dataflows, which are supported by our template.}}

Based on this highly configurable template, we have developed matching delay, \textcolor{black}{energy consumption}, and MC evaluators as introduced in Sections~\ref{ch5_2_3:performance_evaluate} and ~\ref{ch5_3:money_evaluate}, which enable our comprehensive design space exploration.

\section{LP Spatial Mapping Encoding}\label{ch4:encoding}

\subsection{Encoding Format and Parsing Methods}\label{ch4_1:analysis}
\noindent In this section, we introduce a layer-centric encoding method for describing LP SPM schemes. At a high-level, the encoded LP SPM scheme encapsulates two key aspects of information: (1) the allocation and partition of each layer to specific cores for computation, and (2) the data sources and destinations for the workload on each core. This encoding method exhibits great generality and can adapt seamlessly to diverse NoC topologies and core microarchitectures (demonstrated in Sec.~\ref{ch6_2_2:tarch}). Subsequently, we provide a comprehensive description of our encoding format and the corresponding parsing method.

\lp{Consider an $N$-layer DNN DAG that needs to be spatially mapped onto an accelerator in an LP manner. This accelerator features a core group $CG$ comprising $M$ cores and $D$ DRAMs. The layers in the DAG form a Layer Group (LG). For the example in Fig.~\ref{figure:example}, the DAG consists of two Convs ($Layer_{1}$ and $Layer_{2}$), forming a layer group with these two layers.} The accelerator in this example has a core group containing six cores and two DRAMs.


In our encoding format, an \textbf{LP Spatial Mapping Scheme} ($LMS$) of a layer group consists of the \textbf{Mapping Scheme} ($MS$) for each layer within it. The $MS$ of layer $i$ has three attributes: Partition ($Part_{i}=(H_{i}, W_{i}, B_{i}, K_{i})$), Core Group ($CG_{i}=(C_{id_{i,1}},C_{id_{i,2}}, ..., C_{id_{i,nc_{i}-1}})$, where $nc_{i}$ is the number of cores in $CG_{i}$), and Flow of Data ($FD_{i}=(IF_{i}, WGT_{i}, OF_{i})$, $-1\leq IF_{i}, WGT_{i}, OF_{i}\leq D$). The left side of Fig.~\ref{figure:example} shows the $LMS$ of a layer group containing two layers and the $MS$ of each layer.

$Part_{i}$ partitions $layer_{i}$ along each dimension of the four-dimensional output cube into approximately equal $nc_{i}$ parts for each core in $CG_{i}$. As shown in Fig.\ref{figure:example}, the four dimensions are Batch ($B$), representing the number of samples processed at a pipeline stage; ofmaps channel, which is also the weight kernel ($K$); ofmaps Height ($H$); and ofmaps Width ($W$). $B_{i}$, $K_{i}$, $H_{i}$, and $W_{i}$ represent the partition of the corresponding dimension. Based on this ofmaps partition scheme, the partition schemes for ifmaps and weights can be uniquely determined based on the features of different types of layers. For the example in Fig.\ref{figure:example}, $layer_{1}$ is a Conv, with B and K equal to 2 and 2, respectively (other dimensions are not partitioned in this example). The example shows how $Part_{1}$ partitions $layer_{1}'s$ ofmaps (on the left of the arrow) and how the partition scheme of the ofmaps deduces the corresponding ifmaps and weight partition schemes (on the right of the arrow).

$CG_{i}$ contains the cores dedicated to computing $layer_{i}$. $CG_{i}$ is ordered ($(C_{1}, C_{2}) \neq (C_{2}, C_{1})$). Each core of it can be an arbitrary core in $CG$. As shown in Fig.~\ref{figure:example}, $CG_{1}$ is $(2,1,5,4)$.

We then establish a Correspondence Rule to map each partitioned workload ($PW$) of $layer_{i}$ to a corresponding core in $CG_{i}$. First, we assign each $PW$ a unique 4-dimension ID based on its location in the partitioned ofmaps cube $(PW = (h,w,b,k), h \in [0,H_{i}), w \in [0,W_{i}), b \in [0,B_{i}), k \in [0,K_{i})$. Then based on the 4-dimension ID, we transform it into a numerical ID, which equals to $h\times W_{i}\times B_{i}\times K_{i}+w\times B_{i} \times K_{i} +b\times K_{i}+k$. The numerical ID ($NID$) of each partitioned workload corresponds to the $(NID+1)$th core in $CG_{i}$ it is assigned to. For example, in Fig.~\ref{figure:example}, the top left subfigure of the four is $layer_{1}$'s first part of both batch and ofmaps channel, and $H$ and $W$ of $layer_{1}$ are not partitioned; thus, its four-dimension and numerical ID are $(0, 0, 0, 0)$ and $0$, respectively. It is mapped to the first core ($C_{2}$) in $CG_{1}$.


$FD_{i}$ represents the data sources of $layer_{i}$'s ifmaps ($IF_{i}$) and weights ($WGT_{i}$), and the destination of $layer_{i}$'s ofmaps ($OF_{i}$). The flow of data can be categorized into two types: those that require explicit management (non-negative values) and those that do not require explicit management or are directly absent ($-1$). 

The scenarios that require explicit management are as follows: (1) For ofmaps, when the subsequent layer is not in the same layer group as the current layer or when the output of the current layer is the output of the entire DNN, it is necessary to explicitly manage the temporary storage of this layer's output to a specific DRAM. (2) For ifmaps, explicit management is only required when the input of the current layer is the input of the entire DNN; otherwise, the data can be fetched from the DRAM where the previous layer's ofmaps were stored. (3) For weights, explicit management is required whenever a layer has weights. Hence, the primary concern for explicitly managing data flow is determining from which DRAM to store or fetch data. In cases where the value is greater than 0, it represents the ID of the DRAM. Meanwhile, 0 represents a special case - interleaving, where we evenly distribute data across all DRAMs to fully and evenly utilize the available bandwidth. For example, as depicted in Fig.~\ref{figure:example}, due to $layer_{1}$ having the input of the whole DNN and weights, $IF_{1}$ and $WGT_{1}$ should be non-negative. $FD_{1}$ of $layer_{1}$ in the example is (1,1,-1), indicating that the ifmaps and weights of $layer_{1}$ originate from the DRAM 1. 

If both layers sharing dependency are in the same layer group, the ofmaps of the previous layer and the ifmaps of the subsequent layer need not be explicitly manipulated because the destination of each part of the previous layer and the source of each part of the subsequent layer can be directly inferred based on the partition schemes and CGs of these layers. Moreover, for the layers without weights, their $WGT_{i}$ value would be -1. For example, in Fig.~\ref{figure:example}, based on $Part_{1}$, $CG_{1}$, $Part_{2}$, and $CG_{2}$, the data communication dependency between cores of $layer_{1}$ and $layer_{2}$ can be deduced directly. Thus, $OF_{1}$ and $IF_{2}$ are -1.

 \begin{figure*}[!thb]
     \centering
     \includegraphics[width=7in]{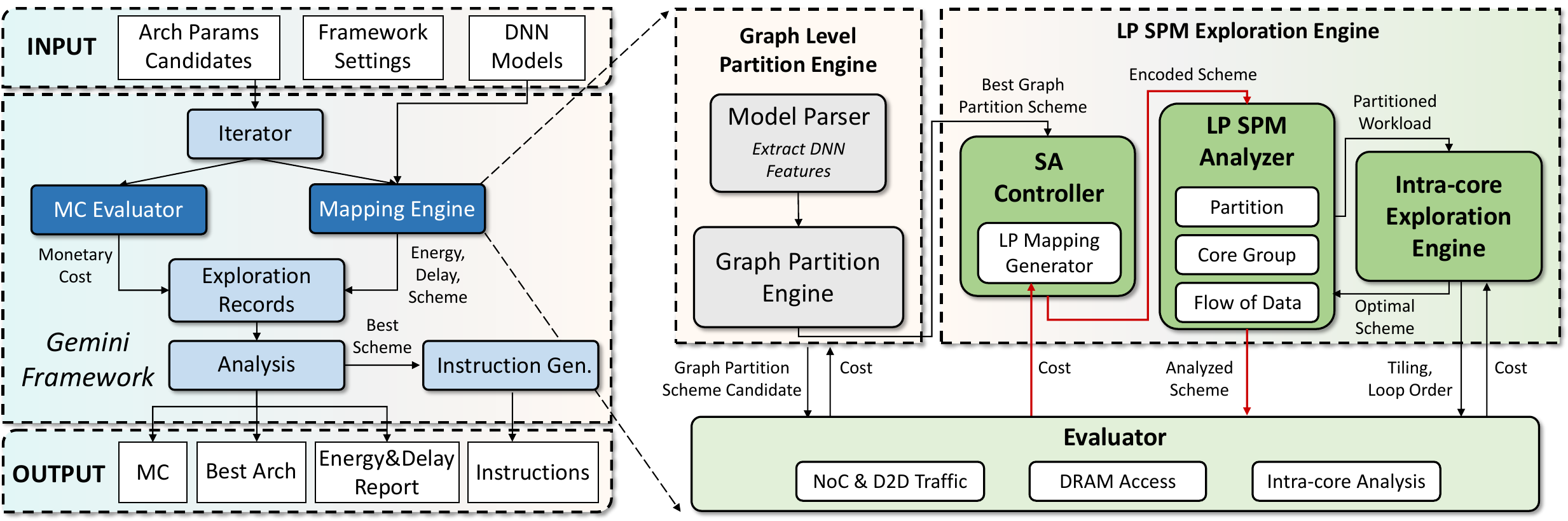}
     \caption{Gemini Framework. The red arrow illustrates the Simulated Annealing (SA) iteration loop.}
     \label{figure:framework}

 \end{figure*}

\subsection{Space Calculation}\label{ch4_3:space_size}
\noindent As shown in Fig.~\ref{figure:example}, each $LMS$ is a point in the optimization space defined by our encoding method. The optimization space of mapping $N$ layers onto an accelerator with a core group containing $M$ cores and $D$ DRAMs is considerably large and extremely complex to calculate. Therefore, we conservatively approximate its lower bound size at $m!\sum\limits_{i=0}^{N-1} {N \choose i}{M-N-1 \choose N-i-1} 4^{N-i}$ schemes, where $x \choose y$ is the binomial coefficient, which equals $\frac{x!}{y!(x-y)!}$. As a comparison, the upper bound of optimization space of the SOTA heuristic Tangram is $N\cdot part(M)$, where  $\textit{Part}_{M}$ represents the total number of possible factorizations for a given integer $M$. We can observe that the optimization space defined by our encoding method is significantly larger than the optimization space of the Tangram heuristic. The detailed calculation procedure and tables of the optimization space under different $M$ and $N$ for our method and Tangram can be found at this link\cite{Space_calculation}.

\subsection{Unveil Hidden \textcolor{black}{Optimization} Opportunities}\label{ch4_4:trade-off}
\noindent In this section, we will introduce the general optimization opportunities for multi-core DNN accelerators hidden within this space, as well as how these opportunities become even more significant in chiplet scenarios.




First, different $Part$ attributes of each layer impact two aspects: (1) NoC communication volume: Various partition schemes lead to different data requirements for each core, causing disparities in NoC communication volumes, even with multicast capabilities. For instance, in Fig.~\ref{figure:example}, under $Part_{1}$, each core requires half of the ifmaps and weight. However, if the $Part$ is changed to $(1,1,1,4)$, each core needs the entire ifmaps and only 1/4 weights. (2) Intra-core optimization space: The dimension of the partitioned workload generated by distinct partition schemes differ, subsequently affecting the optimal intra-core dataflow scheme.

Secondly, the number and position of cores can vary in each $CG$ attribute. The number of cores affects the computation time of each layer, which in turn influences the overall pipeline computation time, as the slowest stage can stall the pipeline. Core positions can significantly impact the data transmission volume and congestion levels of the NoC.

Third, different $FD$ attributes affect the bandwidth utilization and access patterns of different DRAMs, and NoC communication. As shown in Fig.~\ref{figure:example}, the ifmaps and weights of $layer_{1}$ are read from $DRAM1$, while the weight and ofmaps of $layer_{2}$ are read from and written into $DRAM2$, respectively. In this case, the bandwidth demand and access patterns for each DRAM are not balanced, but the total NoC hops are relatively few. If all positive values in $FD_{1}$ and $FD_{2}$ change to 0 (interleaved), DRAM bandwidth usage will become more balanced. However, the total NoC hops increase since some data will interact with more remote DRAM.

\section{Gemini Framework}\label{ch5:gemini}
\subsection{Gemini Overview}\label{ch5_1:overview}
\noindent Gemini is a mapping and architecture co-exploration framework for DNN inference chiplet accelerators. As depicted in the left of Fig.\ref{figure:framework}, the inputs for Gemini consist of (1) Architecture parameter candidates: Each configurable architecture parameter (introduced in Sec.\ref{ch3:hardware_model}) is assigned a list of candidate values; (2) Framework settings: These include the optimization goals and constraints, hyperparameters, and other relevant settings; (3) DNN models: Considering that an accelerator is often used to accelerate different DNNs under different scenarios, Gemini supports conducting DSE for $n$ DNNs.

All architectural candidates are exhaustively explored with the optimization objective of $MC^{\alpha} \times E^{\beta} \times D^{\gamma}$, \textcolor{black}{where MC, E, and D denote Monetary Cost, Energy Consumption, and Delay, and $\alpha$, $\beta$, and $\gamma$ denote the respective importance of MC, E, and D (performance)}. Note that the delay associated with small and large batch sizes can be utilized to illustrate performance in latency-sensitive and throughput-sensitive scenarios, respectively. The $E$ and $D$ are not only influenced by the architecture but also by the specific DNN workloads and their corresponding mapping strategies. Thus, the Mapping Engine employs a dynamic-programming-based graph partition algorithm and an Simulated-Annealing-based (SA-based) LP SPM exploration algorithm to optimize the mapping of the $i$th DNN onto the architectural candidate. \explain{SA algorithm is a widely-used optimization algorithm~\cite{SA}.} This optimization process utilizes $E_{i}^{\beta} \times D_{i}^{\gamma}$ as the optimization goal and yields the evaluation of $E_{i}$ and $D_{i}$ for processing the $i$th DNN with the given architecture parameters. Consequently, the overall Energy Consumption and Delay of the architectural candidate are determined by $(\prod_{i=1}^{n} E_{i})^{\frac{1}{n}}$ and $(\prod_{i=1}^{n} D_{i})^{\frac{1}{n}}$, respectively. In contrast, the Monetary Cost remains unaffected by workloads and mapping strategy, enabling the MC Evaluator to directly assess it based on the architecture parameters.

\subsection{Mapping Engine}\label{ch5_2:mapping_engine}

\noindent As shown in \textcolor{black}{the right part of Fig.~\ref{figure:framework}}, each DNN description file is first processed by the model parser, which generates a DNN topology graph and extracts the features of each layer. Then, this information is sent to the graph partition engine, which partitions the DNN graph into layer groups. Based on the explored graph partition scheme, the LP SPM exploration engine employs an SA-based algorithm with five specifically-designed operators to explore the LP SPM optimization space defined in Sec.~\ref{ch4_3:space_size} for each layer group. 

Since the graph partitioning problem has been extensively studied~\cite{EfficientSchedulingliuleibo,NASA,tangram,atomic,DP_jiazhihao}, we employ the same DP-based graph partition algorithm as Tangram~\cite{tangram}, which is also our baseline, to ensure a fair comparison in our experiments. This algorithm not only generates layer groups efficiently but also determines the number of samples (batch unit) processed in each pipeline stage~\cite{tangram}. 


\subsubsection{LP SPM Exploration Engine}\label{ch5_2_2:LP_explore}
\noindent In this engine, Gemini employs an SA-based algorithm with five specifically-designed operators to explore the broad space defined in Sec.~\ref{ch4:encoding} based on the graph partition scheme explored before.

For each layer group, the initial LP SPM scheme is obtained using a widely adopted heuristic stripe-based strategy~\cite{tangram,atomic,scaledeep}. Then, the SA iteration starts, and the iteration loop is represented by the red arrow in Fig.\ref{figure:framework}. In each iteration, the SA controller randomly selects a layer group with a probability distribution proportional to their optimization size as calculated in Sec.~\ref{ch4_3:space_size}. It then randomly selects one of the five operators to apply a transformation to the chosen layer group. Following this, the LP SPM Analyzer analyzes the modified scheme as introduced in Sec.~\ref{ch4_1:analysis}. After analyzing the partition attribute of each layer, the partitioned workload will be scheduled in intra-core exploration engine, which performs exhaustive search optimization for tiling and loop reorder like many existing works~\cite{understand,timeloop,interstellar,baton,magnet}. Once the optimal solutions of intra-core dataflow for all partitioned layers are found, they will be sent along with the other analyzed information to the Evaluator for overall assessment. If the overall cost is lower, the change is accepted; otherwise, it is accepted with a probability which decreases as the number of iterations increases. 

\vspace{1mm}
\noindent\textbf{SA Operators}: We develop five SA operators to facilitate the exploration process in SA. The operators are as follows:

\noindent\textbf{OP1}: Randomly select a layer and change the values in its $Part$, while still satisfying the constraints of $Part_{i}$.

\noindent\textbf{OP2}: Randomly select a layer and randomly swap two cores within its $CG$, which is equivalent to exchanging the workload between these two cores for a single layer randomly.

\noindent\textbf{OP3}: Randomly select two layers and swap two cores within their $CGs$, which is equivalent to exchanging the workload between these two cores for two layers.

\noindent\textbf{OP4}: Randomly select two layers, remove a core randomly from the $CG$ of one layer, and add it to the $CG$ of the other layer. After the operation, update the $Parts$ of both layers randomly to match their new $CG$ sizes.

\noindent\textbf{OP5}: Randomly select a layer, then choose a non-negative item in its $FD$ randomly, and update its value within the range of 0 to the number of DRAMs randomly.

Utilizing these operators allows each attribute to transition into any other state that fulfills the corresponding constraints through a sequence of transformations. For instance, the size of $CG_{1}$ in Fig.~\ref{figure:example} can be modified to any number from 1 to 5 through a series of $OP4$ operations. Crucially, this ensures that any point within the LP SPM optimization space can be reached from any other \proof{(demonstration link~\cite{Completeness_proof})}, thereby guaranteeing comprehensive exploration by the SA algorithm, and yielding near-optimal solutions.

Through our SA-based algorithm combined with our specially designed operators, Gemini not only can explore the optimization space to balance the trade-offs introduced in Sec~\ref{ch4_4:trade-off}, but also automatically optimize D2D link communication. Since D2D links tend to have smaller bandwidth and higher \textcolor{black}{energy consumption}, during the iterative process, if an SA operation increases the use of more D2D links, it is more likely to significantly reduce performance and energy efficiency, making it less likely to be accepted. On the other hand, SA operations that reduce D2D link usage are more likely to be accepted. Therefore, the entire search process inherently optimizes D2D communication, which will be demonstrated in Sec.~\ref{ch6_4:example}. Furthermore, this mapping technique can aid in architecture design by enabling accelerators to be equipped with lower D2D bandwidth, reducing their area overhead and reaping the benefits of chiplet-induced yield improvement while incurring only minimal performance and energy efficiency losses.


\subsubsection{Evaluator}\label{ch5_2_3:performance_evaluate}
\noindent The Evaluator in the Gemini framework, which is modified based on SET~\cite{SET}, has two interfaces, one for intra-core evaluation and the other for global evaluation. When called by the intra-core exploration engine, the Evaluator can calculate the number of operations for each part, such as the memory access times for different-level buffers and the number of different-precision multiply-accumulate (MAC) operations. Based on this information, the total \textcolor{black}{energy consumption} can be calculated by summing up the number of operations for each component multiplied by the corresponding unit \textcolor{black}{energy consumption}. The overall computation time for the workload on the core can be determined by taking the maximum value among the MAC computation time and the data access amounts for each memory level divided by their respective access bandwidths.

Based on the LP SPM scheme analyzed by the Analyzer and the explored intra-core scheduling scheme for each partitioned layer, the Evaluator conduct the global evaluation by analyzing the data communication volume on each on-chip network link and D2D link, access patterns of DRAM, memory access times for different-level buffers within each core, and operation counts for various-precision computation units. Subsequently, a simulator assesses the delay of the DNN with specific batch size on this accelerator. The \textcolor{black}{energy consumption} can be calculated by summing up the number of operations for each component in the accelerator, multiplied by their respective unit \textcolor{black}{energy consumption}. It is worth mentioning that the \textcolor{black}{energy consumption} of NoC routers is predominantly attributed to the input buffer and crossbar components. Therefore, the per-flit \textcolor{black}{energy consumption} of NoC routers does not vary significantly across different traffic patterns and can be considered a constant value~\cite{orion}. \grs{The \textcolor{black}{energy consumption} calculation for D2D links can be categorized into two types. The first type is for the clock-embedded D2D, where the clock signal does not have a dedicated channel and needs to be recovered from the data. A typical example is SerDes~\cite{serdes2,serdes3,serdes4}. For this type of D2D, almost the same amount of power is consumed regardless of whether data is being transmitted. Therefore, its \textcolor{black}{energy consumption} is calculated as \textit{Number of D2Ds} $\times$ \textit{Power per D2D} $\times$ \textit{Latency}. The second type is the clock-forwarding D2D, which has a separate clock channel, such as GRS~\cite{grs2013,grs2019jssc}, and UCIe~\cite{UCIe}. This type of D2D can enter a low-power state when not transmitting data. Hence, its \textcolor{black}{energy consumption} is calculated as \textit{D2D Communication Volume} $\times$ \textit{Unit D2D Communication Energy Consumption}, similar to on-chip networks. Since the baseline of this work is Simba with GRS links, the second model is also the default model in the experimental to guarantee a fair comparison}. 


\subsection{Monetary Cost Evaluator}\label{ch5_3:money_evaluate}

\noindent MC Evaluator can assess the production cost of different architecture candidates, which mainly includes chiplet \textcolor{black}{silicon} cost, DRAM cost, and packaging cost. 


The evaluation of the silicon area for all chiplets ($Area_tot = \sum_{i=1}^{n} Area_{Die_{i}}$) serves as a fundamental cornerstone within the MC Evaluator. The chip \textcolor{black}{silicon} cost and packaging cost are directly influenced by this area evaluation, as shown in the following paragraphs. Each chiplet's area is determined by the summation of the areas of its constituent modules. In this work, the area of analog IPs, such as PCIe PHY, DDR PHY, and D2D PHY, is obtained directly from the corresponding datasheets. For the logic modules, their area estimation is based on the Verilog code and evaluation process employed during the development of our chip.


The silicon cost of a die (chiplet) is $Area_{Die_{i}} / Yield_{die_{i}} \cdot C_{silicon}$, where $Yield_{Die_{i}} = Yield_{unit}^{Area_{Die_{i}}/Area_{Die_{unit}}}$~\cite{chiplet_yinxiao}. $Yield_{unit}$ represents the yield per unit area ($Area_{Die_{unit}}$), where $Yield_{unit}$ under 12nm in this paper is assumed to be 0.9, and $Area_{Die_{unit}}$ is set to 40 $mm^2$.




The DRAM cost is $\lceil DRAM_{bw}/Unit_{bw}\rceil \cdot C_{DRAM_{Die}}$, where $Unit_{bw}$ and $C_{DRAM_{Die}}$ represent the bandwidth provided by each DRAM die and the MC of each DRAM die, respectively. The $Unit_{bw}$ and $C_{DRAM_{Die}}$ used in this work (GDDR6) are 32$GB/s$ and \explain{\textdollar$3.5$~\cite{DRAMPRICE}}.

The packaging cost equals $(Area_{tot} \cdot f_{scale})/Yield_{package} \cdot C_{package}$. Because the substrate requires a larger area than the total area of all chiplets to accommodate IO fanout and interconnect wiring functions, there is an empirical scaling factor $f_{scale}$ to calculate the substrate area based on total silicon area~\cite{chiplet_yinxiao}. $C_{package}$ represents the monetary cost per unit area of the substrate. The $C_{package}$ for different substrates varies. For instance, in the case of an organic substrate, without adopting chiplet technology, a standard fan-out substrate is relatively inexpensive (0.005\textdollar$/mm^{2}$). However, when chiplet technology is employed, the price increases due to the need to support high-density interconnects. $C_{package}$ varies across different area ranges. Larger substrate areas require more intricate manufacturing processes, resulting in higher $C_{\text{package}}$.

\subsection{\future{Future Work Grounded on Gemini}}\label{ch5_5:future}
\noindent The field of DNN chiplet accelerators is still emerging, with limited research available. Gemini offers a fertile ground for further exploration. In this section, we outline examples of promising future work in the realms of mapping and architectural design that can be pursued using Gemini. Our intention is to underscore the value of Gemini while also fostering community development.

At the mapping level, a promising direction is to co-explore the SPM optimization dimension with other optimization dimensions, such as the graph-level composite spatial-temporal dimension defined by SET. Jointly exploring these dimensions could potentially yield better solutions, and it would also be valuable to investigate the interplay between these different optimization dimensions On the architectural front, the heterogeneity of chiplet presents another compelling area for research. Questions around scheduling LP mapping on heterogeneous chiplets and, reciprocally, exploring architectural designs for heterogeneous accelerators in the context of LP mapping are of particular interest.

\section{Evaluation}\label{ch5:evaluation}

\subsection{Experiment Setup}\label{ch5_1:setup}

\begin{table}[!b]
\centering

\caption{DSE Parameters. The parameters marked in \underline{\textit{italics}} and bold are exclusively used by DSE for 72TOPs and 128/512 TOPs accelerators, respectively.}\label{Table1:DSE}
\renewcommand{\arraystretch}{1.3}
\begin{tabular}{|c|c|c|}
\hline
\multirow{3}{*}{\textbf{\begin{tabular}[c]{@{}c@{}}Accelerator \\ Config.\end{tabular}}} & $X_{Cut}$          & \underline{\textit{1, 2, 3, 6}} (\textbf{1, 2, 4, 8})                                                                                            \\ \cline{2-3}  & $Y_{Cut}$          & \underline{\textit{1, 2, 3, 6}} (\textbf{1, 2, 4, 8})     \\ \cline{2-3}  & $DRAM_{BW}$   & 0.5, 1, 2 GB/s per TOPs        \\ \hline
\multirow{2}{*}{\textbf{\begin{tabular}[c]{@{}c@{}}Network \\ Config.\end{tabular}}}     & $NoC_{BW}$         & \underline{\textit{8}}, 16 ,32, 64, \textbf{128} GB/s                                                                               \\ \cline{2-3}  & $D2D_{BW}$         & NoC/4, NoC/2, NoC           \\ \hline
\multirow{2}{*}{\textbf{\begin{tabular}[c]{@{}c@{}}Core \\ Config.\end{tabular}}}        & $GBUF/Core$            & \multicolumn{1}{c|}{\begin{tabular}[c]{@{}c@{}} 256, 512, 1024, 2048, \\ 4096, \textbf{8192 KB}\end{tabular}} \\ \cline{2-3}  & $MAC/Core$ & \underline{\textit{512}}, 1024, 2048, 4096, \textbf{8192}                                                                               \\ \hline
\end{tabular}
\end{table}

\subsubsection{DSE Configurations}\label{ch5_1_2:hardware_config}
\noindent In this work, we conduct DSE on DNN accelerators with 72 TOPs, 128 TOPs, and 512 TOPs to show Gemini advantages and gain a deeper understanding of the design space for DNN chiplet accelerators and the use of chiplet technology. We choose 72 TOPs instead of 64 TOPs to enable comparison with the SOTA Simba architecture with 72 TOPs~\cite{simba}. Table~\ref{Table1:DSE} presents the DSE parameters. By keeping the total computing power constant, we determine the number of cores based on the MAC/Core configurations. To maintain the core array's length and width as close as possible, we arrange the cores accordingly. For instance, with 36 cores, we configure them in a 6$\times$6 arrangement, while for 18 cores, we adopt a 6$\times$3 configuration. The values of $X_{Cut}$ and $Y_{Cut}$ must be a factor of the corresponding number of cores on edge; otherwise, the candidate is deemed invalid.


We choose TSMC 12nm as the default process, which is also widely used by many commercial accelerators~\cite{hanguang,tpuv2,huawei}. We choose organic substrate as the default packaging substrate, as it is used in many successful commercial~\cite{zen2,zen3,zen4} and academic designs~\cite{simba}. The default operating frequency is set to 1GHz. For D2D, we use GRS~\cite{simba,grs2019jssc} as default in experiments to guarantee a fair comparison with Simba.


\textcolor{black}{For the sake of brevity, the explored architectural parameters are presented as (\textit{Chiplet Number, Core Number}, $DRAM_{BW}, \\NoC_{BW}, D2D_{BW}, GBUF/Core, MAC/Core$).}



The DSE adopts a batch size of 64, targeting a throughput-driven scenario, as defined by the MLPerf benchmark~\cite{mlperf}. Although the exploration process solely utilizes these two DNN models with a batch size of 64, we also evaluate additional DNNs with batch size 1 (latency-centric scenarios) on the explored architecture, demonstrating its broad applicability to other scenarios. We set the default optimization objective of DSE as $MC\cdot E\cdot D$ (as introduced in Sec.~\ref{ch5_1:overview}). The default workload is set as Transformer~\cite{transformer} since it is prevalent and widely used in various scenarios such as image~\cite{vision_transformer} and language processing~\cite{gpt-3,gpt4,bert}. 

\subsubsection{DSE Time} The DSE time increases with the increase in target computing power. For example, the DSE for 72 TOPs and 512 TOPs use 80 threads and 100 threads, respectively, and run for 2280s and 23907s on an Intel Xeon Platinum 8260 server.


\subsubsection{Workloads}\label{ch5_1_1:workloads}
\noindent To fully demonstrate the advantages of our architecture and mapping co-exploration, we conduct the comparison with the baseline over a wide range of DNNs, including ResNet-50 (RN-50)~\cite{Resnet}, ResNeXt (RNX)~\cite{resnetxt}, Inception-ResNet (IRes)~\cite{inception-v4}, PNasNet (PNas)~\cite{Pnas}, and Transformer (TF)~\cite{transformer}. ResNet-50 and ResNeXt are selected due to their use of classic residual structures, which are prevalent in many DNNs. Inception-ResNet-v1 and PNASNet represent DNNs with more intricate dependencies. Due to the reasons mentioned above, we mainly use Transformer as a representative to showcase various observations and insights, in addition to overall comparisons.


\subsubsection{Baseline}\label{ch5_1_3:baseline}
\noindent The baseline here principally comprises two components: baseline architecture and baseline mapping. For the architecture baseline, we adopt Simba~\cite{simba} and optimize it accordingly. Since the Simba accelerator is a test chip lacking DRAM, the configuration of its on-chip buffer is also impacted. Consequently, we initially equip it with I/O Dies and furnish a total DRAM bandwidth of 2 GB/s per TOPs. Concurrently, the on-chip buffer configuration is determined according to the Simba-series paper~\cite{magnet}, which explores the architecture of an NVDLA-style core. In particular, the GBUF is allocated 1024 KB per core. The remaining configurations align with the specifications in the original Simba paper~\cite{simba}. In experiments, the baseline architecture is abbreviated as \textbf{S-Arch}, while the architecture explored by our Gemini framework is denoted as \textbf{G-Arch}. Additionally, to demonstrate the universality of Gemini, we conduct an exploration with a hardware template modified to adopt a folded torus NoC topology. We compare the explored architecture and mapping scheme with an accelerator that employs the Tenstorrent Grayskull architecture parameters~\cite{tenstorrent} (\textbf{T-Arch}) and utilizes Tangram mapping. This comparison serves to validate the effectiveness of the Gemini (Sec.~\ref{ch6_2_2:tarch}).


Although Simba has its own layer-sequential mapping~\cite{simba}, this mapping performs poorly on such a high-computing-power chip~\cite{tangram,atomic}. Therefore, we choose the SOTA Tangram LP mapping strategy as the baseline mapping strategy (abbr., \textbf{T-Map}). The mapping scheme explored by Gemini framework is abbreviated as \textbf{G-Map}. 
\begin{figure}[!thb]
    \centering
    \includegraphics[width=3.2in]{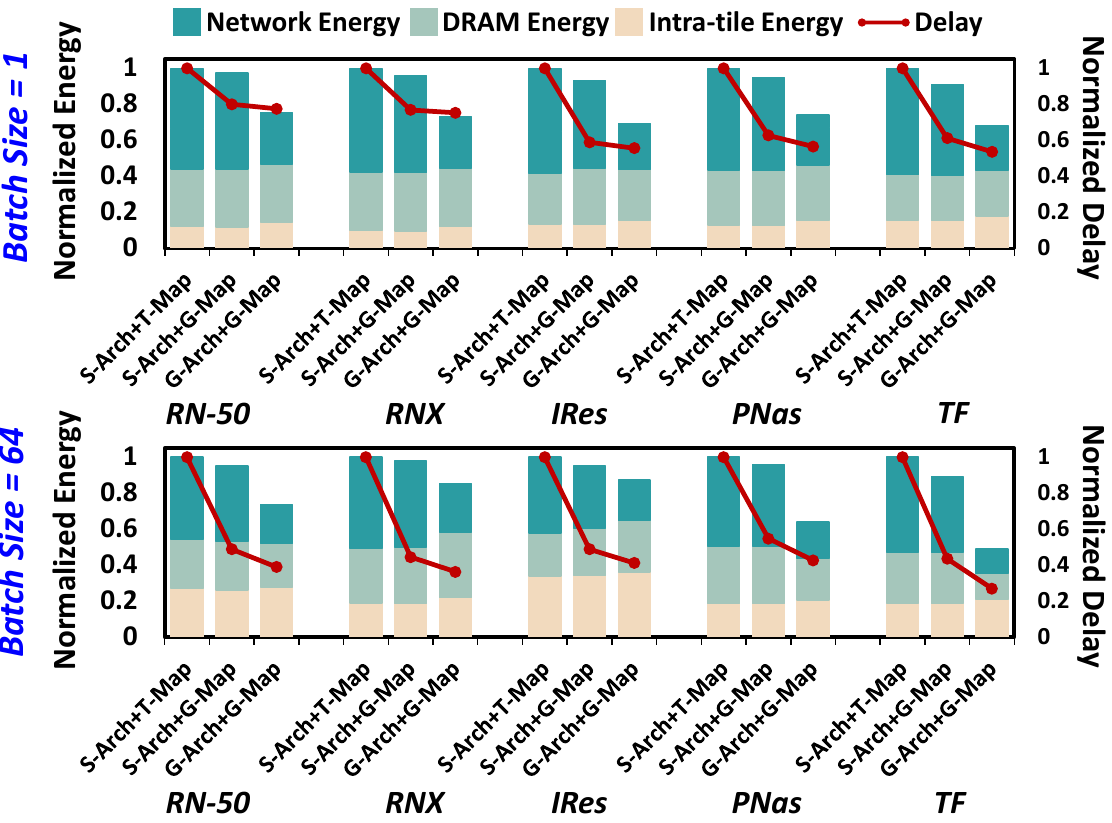}
    \caption{Overall Comparisons among G-Arch, G-Map, S-Arch, and T-Map. \textcolor{black}{energy consumption} and delay in each comparison are normalized to overall baseline \textbf{S-Arch+T-Map}}
    \label{figure:overall_compare}

\end{figure}
\subsection{Overall Comparison}\label{ch6_2:overall_compare}
\subsubsection{Compared to S-Arch with T-Map} As illustrated in Fig.~\ref{figure:overall_compare}, G-Arch mapped with G-Map achieves a 46.8\% reduction in delay and a 28.8\% reduction in \textcolor{black}{energy consumption} across five DNNs and two batch sizes when compared to S-Arch mapped with T-Map. This improvement is attained with only a 14.3\% additional MC, demonstrating the value of Gemini co-exploration. Moreover, by merely employing G-Map on the S-Arch, a significant reduction delay and energy consumption can be achieved compared to T-Map. The optimization brought up by Gemini mapping demonstrates that exploring the vast space defined by our encoding method can indeed achieve a better balance of the trade-offs introduced in Sec.~\ref{ch4_4:trade-off}. \sgarch{The explored G-Arch is (2, 36, 144GB/s, 32GB/s, 16GB/s, 2MB, 1024). Compared to S-Arch, the number of Chiplets in G-Arch is significantly reduced, while the bandwidths of NoC and D2D are increased, and the GBUF capacity is doubled. The reduction in the number of Chiplets leads to a significant decrease in the proportion of D2D links across all links, resulting in a substantial reduction in network energy. With improved interconnect bandwidth and on-chip storage resources, performance and energy efficiency improvements are natural. However, the interesting point is that these resource improvements only result in a 14.3\% increase in MC. This is mainly because, under S-Arch, there are too many Chiplets, and an excessive amount of chip area is used for D2D interfaces (nearly 40\%). In G-Arch, this area is converted into interconnect bandwidth and on-chip storage resources. However, it is worth noting that Simba serves as an academic test chip. As a pioneering work in introducing Chiplets into DNN accelerators, Simba has already showcased many features of Chiplet-based inference accelerators. Given that Simba is an academic demo with limited chiplet area and scale (6 $mm^2$~\cite{simba-jssc}), its suboptimal chiplet granularity is understandable. Further insights regarding the optimal choice of chiplet granularity will be discussed in Section~\ref{ch7_1_1:chipletinsights}.}


\subsubsection{Compared to T-Arch with T-Map}\label{ch6_2_2:tarch} We compare the architecture and mapping scheme explored by Gemini with a 120-core monolithic accelerator that utilizes the same architectural parameters as Tenstorrent Grayskull (Core array size, MAC \& GBUF per core, and DRAM bandwidth)~\cite{tenstorrent} and employs Tangram mapping. Gemini's explored architecture is (6, 60, 480 GB/s, 64GB/s, 32GB/s, 2MB, 2048). Compare with T-Arch with T-Map, G-Arch with G-Map achieves $1.74 \times$ performance and $1.13 \times$ energy efficiency, and reduces $40.1$\% MC, which demonstrates the effects and universality of Gemini.

\section{Discussion}\label{ch7:discussion}

\subsection{Design Space Exploration}\label{ch6_3:DSE}

\subsubsection{\chipletg{Chiplet Granularity}}\label{ch7_1_1:chipletinsights}
\begin{figure}[!t]

     \centering
     \includegraphics[width=3.3in]{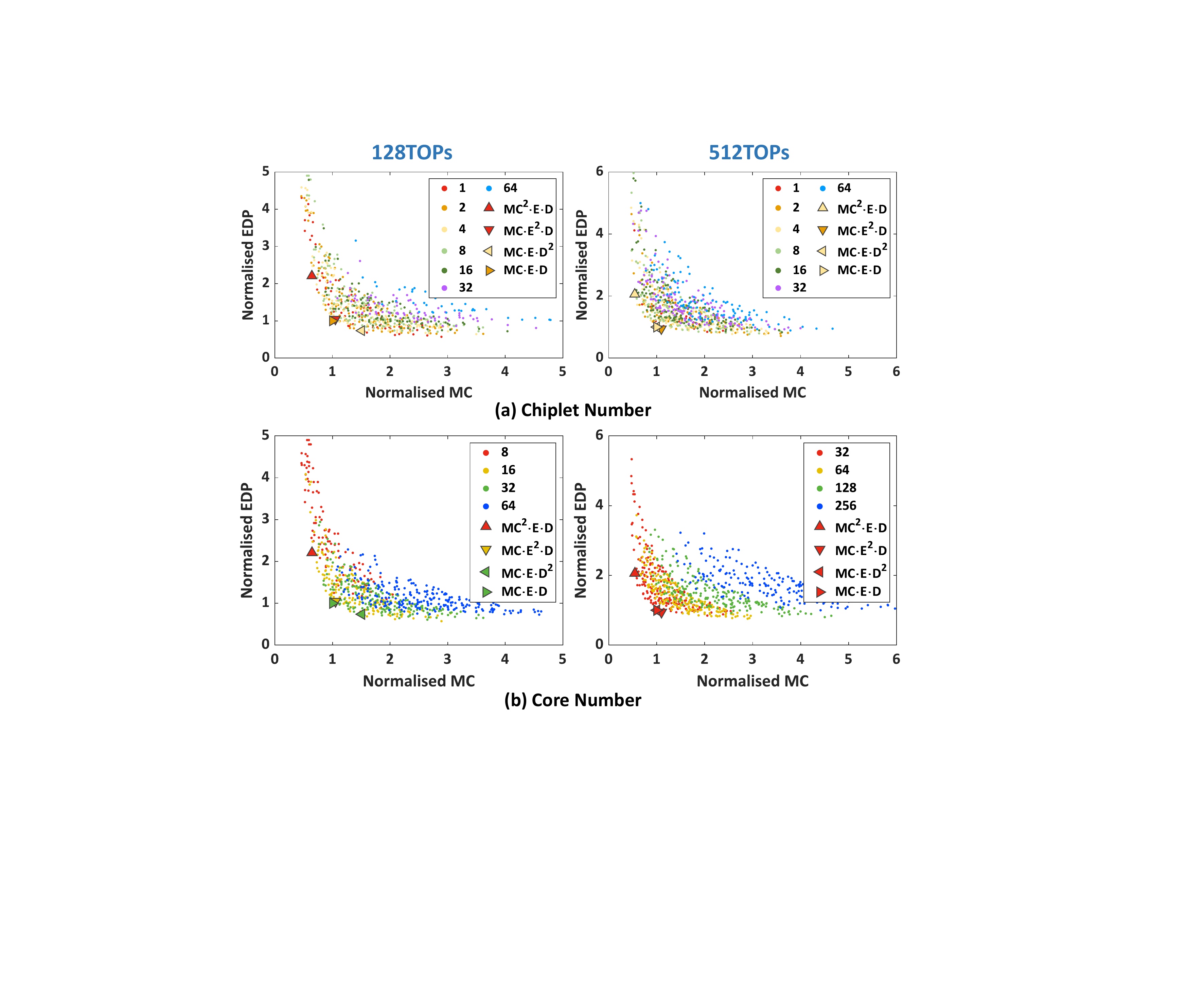}

     \caption{\coreg{EDP and MC of the Architecture Candidates in the Design Space for 128 and 512 TOPs Accelerators. Different colors represent various categories of architectural parameters. To provide a clearer depiction of trends, only the top 50\% of each category is plotted. The EDP and MC of each point are normalized to the counterpart of the best arch under $MC\cdot E \cdot D$. We use triangles pointing in different directions to represent the globally optimal architectures under different optimization objectives. The workload is Transformer.}}
     \label{figure:DSE}

 \end{figure}

\noindent \chipletg{As illustrated in Fig.~\ref{figure:DSE}(a), the optimal number of chiplets for 128 TOPs and 512 TOPs, under the four different objectives, ranges from 1 to 4 and 2 to 4, respectively. Moreover, it is evident from both DSEs that an excessively fine-grained chiplet granularity negatively impacts MC, energy consumption, and performance. Combining it with the analysis in comparing G-Arch and S-Arch in Sec.~\ref{ch6_2:overall_compare}, we can achieve an insight: \textbf{\textit{Partitioning DNN accelerators into chiplets with moderate granularity can effectively strike a balance among MC, energy consumption, and performance. Conversely, an excessively fine-grained approach to chiplet partitioning can have a negative impact on all three metrics—performance, energy consumption, and MC—simultaneously.}}} 

  \begin{figure}[!b]

     \centering
     \includegraphics[width=3.3in]{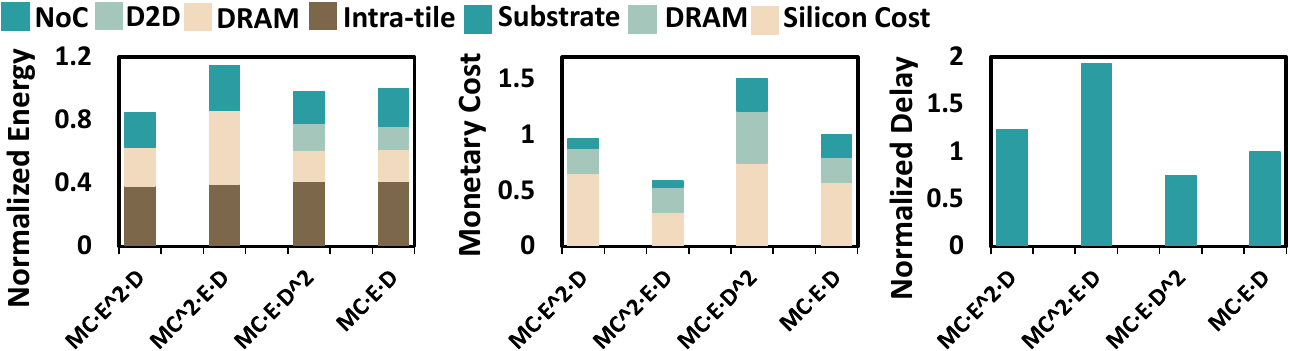}    
     \caption{\coreg{Comparative Analysis of Energy Consumption, MC, and Delay for Optimal Architectures Explored under Four Different Optimization Objectives. In the left-to-right order, they are: (1, 16, 128GB/s, 32GB/s, None, 4MB, 4096), (1, 8, 128GB/s, 32GB/s, None, 4MB, 8192), (4, 32, 256GB/s, 64GB/s, 32GB/s, 2MB, 2048), and (2, 32, 128GB/s, 32GB/s, 16GB/s, 2MB, 2048). All data are normalized to $MC\cdot E\cdot D$.}}
     \label{figure:fourexamples}

 \end{figure}
\subsubsection{\coreg{Core Granularity}}\label{ch7_1_2:coreinsights}
 \coreg{In addition to the granularity of chiplets, the granularity of cores is also an important topic in many-core architectures. Particularly in the context of LP mapping, the trade-offs associated with varying core granularities have yet to be explored. This section aims to address this gap in the literature.}

 \coreg{As shown in Fig.~\ref{figure:DSE}(b), there is a clear trend of increasing MC as the number of cores increases. This is primarily because, to improve overall performance and energy efficiency, it is necessary to increase network bandwidth or buffer capacity per core. The more cores there are, the more these resources multiply, leading to greater area overhead. However, from the perspective of EDP, it initially decreases with an increase in the number of cores, then rises again later. For example, in the 128 TOPs scenario, the EDP for the yellow and green points is better than that for the red and blue points. This intriguing phenomenon warrants further analysis to understand the underlying trade-offs.}

 \coreg{A main advantage of LP Mapping is to reduce DRAM accesses, which is a key factor contributing to improvements in both energy consumption and performance. Under LP Mapping, an increased number of cores facilitates longer pipelines, allowing for the simultaneous processing of more layers. This, in turn, enables more dependency data to be transferred and processed on-chip, thereby reducing costly DRAM accesses. However, not all dependencies have the same amount of data. Layers involving links with larger data volumes will be prioritized for simultaneous processing. Therefore, the reduction in DRAM access due to the increase in the number of layers processed simultaneously diminishes at a slower rate. Moreover, as the pipeline depth increases, so does the utilization loss due to the filling and draining phases~\cite{tangram,atomic}. In summary, a longer pipeline is not necessarily better. The diminishing benefits from extending the pipeline and the escalating costs eventually reach a point of equilibrium, which represents the optimal pipeline length. As evidence, in Fig.~\ref{figure:fourexamples}, the number of cores for the optimal solutions under the four different objectives, from left to right, are 16, 8, 32, and 32. Correspondingly, the average number of layers processed simultaneously for each are 5.4, 4.1, 10.2, and 8.1. As shown in Fig.~\ref{figure:fourexamples} Left, though DRAM access continuously decreases as the number of cores increases from 8 to 32, the rate of reduction is much faster when transitioning from 8 cores to 16 cores (48.0\%) compared to the transition from 16 to 32 cores (average 19.4\%). Moreover, the average number of layers processed simultaneously by the optimal 64-core architecture (the best among all 64-core configurations under the target of $MC\cdot E\cdot D$) is 9 layers, which does not continue to increase with the number of cores, indicating the striking of the balance.}

 \subsection{\SCMS{Reuse A Single Chiplet for Multiple Accelerators}}\label{SCMS}

\begin{figure}[!b]

      \centering
      \includegraphics[width=3.2in]{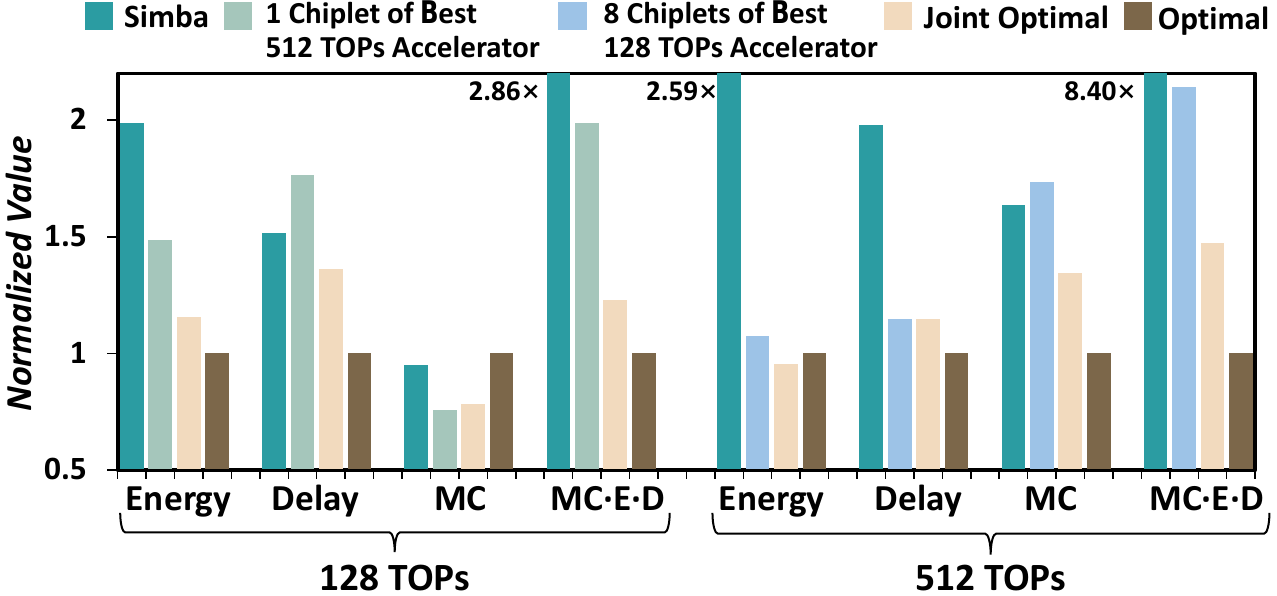}
    
     \caption{\SCMS{Energy, Delay, and MC of Four Construction Schemes for 128 TOPs and 512 TOPs Accelerators. These construction schemes are: constructed by Simba chiplets (\textbf{Simba}), constructed by the chiplets of the optimal explored architecture of the other computing power (\textbf{1 Chiplet of Best 512 TOPs Accelerator} and \textbf{8 Chiplets of Best 128 TOPs Accelerator}), constructed by the optimal solution found through joint architecture exploration of the two computing power (\textbf{Joint Optimal}), and constructed by the optimal architecture explored under the target of its own computing power (\textbf{Optimal}). It is worth noting that the second scheme is feasible because, regrading $MC\cdot E\cdot D$, the optimal architecture for 512 TOPs and 128 TOPs happens to be 4 chiplets and 2 Chiplets (Fig.~\ref{figure:DSE}(b)), so one chiplet of 512 TOPs and 8 Chiplets of 128 TOPs can be used to construct a 128 TOPs accelerator and 512 TOPs accelerator, respectively.}}
      \label{figure:SCMS}

  \end{figure}
 \noindent \SCMS{One significant advantage of utilizing chiplets in the industry lies in their reusability across accelerators of varying scales. This approach has already been successfully employed in CPUs~\cite{zen2,zen3,zen4}. Despite its promise, this potential benefit has yet to be explored within the realm of DNN inference accelerators. This section aims to bridge this gap using Gemini Framework.} 



\SCMS{Figure~\ref{figure:SCMS} reveals that the performance of accelerators constructed with Simba chiplets is poor for both 128 TOPs and 512 TOPs configurations, with the latter faring even worse. \textbf{\textit{This example illustrates the failure of a ``one-size-fits-all'' approach, highlighting the impracticality of designing a small-scale chiplet to cover an extensively wide range of scenarios}}. Additionally, although using chiplets from the best 128 TOPs accelerator in the 512 TOPs accelerator yields better results than Simba, the overall performance is still unsatisfactory, and vice versa. This example, in conjunction with the Simba case, collectively illustrates that \textbf{\textit{directly repurposing chiplets from one computing platform to build another is ill-advised and frequently results in less-than-ideal outcomes}}.} 
  

\SCMS{To fully harness the potential of ``Chiplet Reuse for Multiple Accelerators'', we have enhanced Gemini to enable DSE for multiple computational power levels concurrently. In this approach, Gemini strategically organizes the chiplets of each architecture candidate with the lowest computational power into accelerators designed for higher computational power requirements. Subsequently, the product of $MC\cdot E\cdot D$ for all accelerators is calculated, and the architecture yielding the minimum product is chosen as the optimal solution (Joint Optimal in Fig.~\ref{figure:SCMS}).}

\SCMS{By comparing the Joint Optimal with the individual Optimal, we find that although the Joint Optimal still has a certain gap (with $MC\cdot E\cdot D$ being on average 34\% higher), this gap is much smaller than those of the previous two approaches. This modest overhead is acceptable when considering that chiplet reuse can significantly reduce NRE costs, including one-time expenses such as design, verification, IP acquisition, and tape-out. The benefits of this approach become increasingly significant in advanced process nodes or fragmented markets since NRE costs tend to grow non-linearly with process advancement, while the fragmented market—with its smaller volumes for each market—makes it challenging to amortize expensive NRE costs effectively. Thus, it concludes that \textbf{\textit{with the proper design considerations—facilitated by the support of Gemini—the idea of employing a single chiplet for multiple accelerators can be effectively applied to DNN inference accelerators.}}}

\begin{figure}[!b]
 \vspace{-3mm}
    \centering
    \includegraphics[width=3.2in]{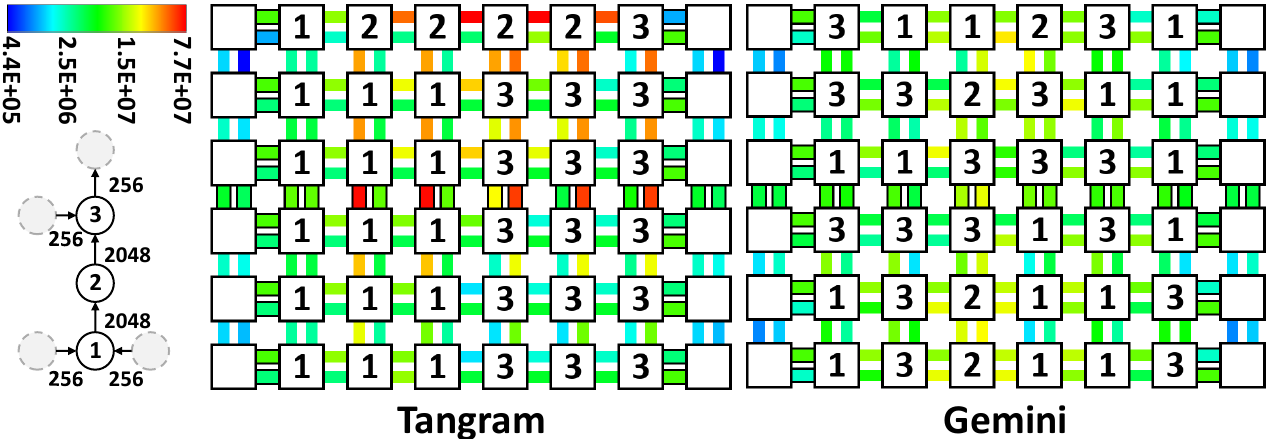}
    \caption{Network Traffic Heatmap of the Optimal SPM Scheme Explored by Tangram and Gemini on 72 TOPs G-Arch. The nodes without numbers at the edges of the heatmap represent communication nodes within the controllers in the IO Die (Fig.~\ref{figure:hardware}(a)). All D2D links in the figure are marked with black borders. Due to the D2D bandwidth being half of the on-chip NoC link, we double the data volume on it to display the bandwidth pressure more clearly. The bottom-left corner depicts the topology and data volume of each dependency of the three-layer workload in Transformer. Layers without numbers represent layers belonging to different layer groups.}
    \label{figure:network}

\end{figure}

\subsection{Learn From an Actual SPM Example}\label{ch6_4:example}

\noindent In this section, we demonstrate the advantages of Gemini mapping through a practical example (Fig.~\ref{figure:network}) and analyze how to allocate computing resources to each layer in a multi-core chiplet accelerator. 


As can be observed in Fig.\ref{figure:network} bottom left, the data volume from $layer_{1}$ to $layer_{2}$ and from $layer_{2}$ to $layer_{3}$ is significantly higher than that of other dependencies. The links with the most data (reddest) in the Tangram SPM scheme are also caused by these two dependencies. For instance, since the data calculated in $layer_{2}$ needs to be sent to $layer_{3}$, under XY-routing, the data calculated by the left cores of $layer_{2}$ is first sent to the right and then downwards, as illustrated in the upper part of the mesh in Fig.\ref{figure:network}. The accumulation of ofmaps data computed by multiple cores of $layer_{2}$ leads to a large amount of communication data on the upper on-chip links. Thus the links are red. During the downward transmission process, each core of $layer_{3}$ consumes a portion of the ofmaps data calculated by $layer_{2}$, causing the actual data volume to decrease gradually. This is reflected by the colour of the downward links becoming lighter (except for the D2D links). However, due to the smaller bandwidth of the D2D links compared to the on-chip links, the bandwidth pressure on the D2D links is higher.

By automatically exploring the vast optimization space we have defined, Gemini mapping can discover solutions that are difficult to design manually. In detail, as shown in Gemini scheme in Fig.~\ref{figure:network}, we can see that the red and orange links have completely disappeared. This is mainly due to two factors: (1) the total hop count decreases by 34.2\%, with a 74\% reduction in hop count on the intermediate D2D links; and (2) Gemini can utilize the originally relatively idle links (blue) more effectively, which is reflected in the decreased number of blue links in the figure. As a result, the overall network traffic is more evenly distributed, significantly improving network performance and overall performance.

From the actual example in Fig.~\ref{figure:network} and other similar instances that are not shown, we can derive an \textbf{insight}: \textit{\textbf{the widely used clustered core allocation strategy~\cite{tenstorrent,simba,tangram,atomic,scaledeep,SET}, where a layer is assigned to consecutive and rectangle-like core group, may not always be a good choice. This is because the dependencies between certain layers may have particularly high data transfer requirements, and clustering cores for each layer together cannot disperse this transfer demand, leading to network congestion.}} 

\section{Conclusion}

\noindent In this work, we introduce Gemini, a mapping and architecture co-exploration framework for large-scale DNN chiplet accelerators, which considers MC, performance, and energy efficiency. For mapping, Gemini employs a novel encoding method to define LP spatial mapping schemes and the corresponding parsing method. This approach enables us to identify the optimization space of LP SPM and unearth hidden optimization opportunities. Additionally, we devise a specially-tailored SA algorithm to efficiently navigate this space. Regarding architecture, we provide a highly configurable hardware template and precise evaluators for performance, energy, and MC. Our experiments demonstrate that Gemini's explored architecture and mapping scheme significantly outperform the SOTA Simba architecture with Tangram mapping, with only a slight increase in monetary cost. We also present insightful observations about the use of chiplet technology in architecture design and DNN workload mapping within chiplet contexts.



  \section*{Acknowledgment}
This research was partially supported by National Key R\&D Program of China (2022YFB2804103), Tsinghua University Dushi Program, and Tsinghua University Talent Program, National Natural Science Foundation of China (62072262).
 

\bibliographystyle{IEEEtranS}
\bibliography{refs}
%
%
%
%
%


\appendix
\section{Artifact Appendix}

\subsection{Abstract}

\noindent This appendix provides guidance on accessing and using the Gemini framework (introduced in Sec.~\ref{ch5:gemini}) to replicate the key results shown in Fig.~\ref{figure:overall_compare}. The process is divided into two steps: (1) Conducting DSE for a 72 TOPs setup (Table.~\ref{Table1:DSE}) to identify the optimal architecture, referred to as G-ARCH; (2) Comparing the efficacy of G-ARCH with G-MAP against the baseline architecture S-ARCH and baseline mapping T-MAP, across varying batch sizes (introduced in Sec.~\ref{ch5_1:setup}) and five different networks (introduced in Sec.~\ref{ch5_1_1:workloads}). The remaining experiments, which also involve similar DSE and analysis, are omitted here for the sake of brevity.
\vspace{-1mm}

\subsection{Artifact check-list (meta-information)}


{\small
\begin{itemize}
  \item {\bf Algorithm: Simulated Annealing, Exhaustive Search and Dynamic Programming}
  \item {\bf Program: C++, Shell, Python (only for data collection)}
  \item {\bf Compilation: by Makefile}
  \item {\bf Hardware: It is best to have a server with more than 80 threads.}
  \item {\bf Metrics: Cost function ($MC^{\alpha} \times E^{\beta} \times D^{\gamma}$, $MC\times E\times D$ is employed in DSE. $E\times D$ is employed in the comparisons with baseline.)}
  \item {\bf Experiments: reproduce Fig.~\ref{figure:overall_compare}, including 72TOPs DSE and comparisons with baseline.}
  \item {\bf How much disk space required (approximately)?: 1GB}
  \item {\bf How much time is needed to prepare workflow(approximately)?: Several minutes at most.}
  \item {\bf How much time is needed to complete experiments (approximately)?: For DSE, it takes about 38 minutes with 80 threads; For comparison, it takes about 14 minutes using 10 threads. The server is Intel Xeon Platinum 8260
server}
  \item {\bf Publicly available?: Yes}
  \item {\bf Code licenses (if publicly available)?: BSD 3-Clause License
}
  \item {\bf Archived (provide DOI)?: 10.5281/zenodo.10207613}
\end{itemize}
}

\subsection{Description}

\subsubsection{How to access}

The artifact is uploaded to Zenodo:10.5281/zenodo.10207613


\subsubsection{Software dependencies}

A C++ compilation environment with support for the C++ 17 standard is required. Linux is recommended. It is recommended to use ``GNU make'' to build the program. Additionally, two python packages ``pandas'' and ``csv'' are needed.



\subsection{Installation}

For artifact evaluation, start by downloading the artifact from Zenodo:

\lstset{basicstyle=\ttfamily\footnotesize,
  breaklines=true,
}

\begin{lstlisting}[]
  $ wget -O GEMINI_AE.zip https://zenodo.org/records/10207613/files/GEMINI_AE.zip?download=1
  $ unzip GEMINI_AE.zip

\end{lstlisting}

Our Gemini exploration framework is in ``GEMINI''.
We use Makefile to build the GEMINI framework.

\begin{lstlisting}[]
  $ cd GEMINI
  $ make
\end{lstlisting}
Then the executable target will be generated at ``./build/stschedule''.

You can install the needed python packages using pip with following commands.
\begin{lstlisting}[]
  $ pip install -r requirements.txt
\end{lstlisting}
Or you can use conda to install with following commands.
\begin{lstlisting}[]
  $ conda install --file requirements.txt
\end{lstlisting}


\subsection{Experiment workflow}

\subsubsection{72TOPs DSE}
Once GEMINI framework is built, you can execute DSE experiment and search the best arch with command below.

\begin{lstlisting}[]
  $ ./dse.sh
\end{lstlisting}
The DSE script uses 80 threads and takes around 38 minutes to run on Intel Xeon Platinum 8260 server.
Then you can see the state of current running DSE. Once the process is completed, the command-line window will output the optimal architecture. You can compare this optimal architecture with the one mentioned in Sec.~\ref{ch6_2:overall_compare}, which is (2, 36, 144GB/s, 32GB/s, 16GB/s, 2MB, 1024). In the "dse\_log" folder, there will be a folder named with the timestamp, containing the outputs for each architecture candidate, as well as the summarized ``result.csv'' file. You can compare the generated "result.csv" with the ``DSE\_result.csv'' provided in "expected\_results" folder, which is employed in our paper.

\subsubsection{Comparison with Baselines}
Once you obtain the optimal architecture, you can proceed with the comparison of G-ARCH+G-MAP, S-ARCH+T-MAP, and S-ARCH+G-MAP for the five networks and two batch sizes to reproduce Fig.~\ref{figure:overall_compare} with following commands.
\begin{lstlisting}[]
  $ ./compare.sh <output_dir>/best_arch.txt
\end{lstlisting}
Please note that ``\textless{}output\_dir\textgreater{}'' refers to the address where the ``best\_arch.txt'' file is stored, which is the folder named with a timestamp within the ``DSE\_log'' directory like ./dse\_log/2023\_11\_08\_10\_56\_52/best\_arch.txt. 

The script uses 10 threads and takes around 14 minutes to run on Intel Xeon Platinum 8260 server. Then you can see the output information in the terminal window about the optimization rate of G-ARCH+G-MAP over S-ARCH+T-MAP, which will be identical with the results posted in Abstract and Sec.~\ref{ch6_2:overall_compare}(1.98× performance improvement, 1.41× energy
efficiency improvement, and 14.3\% MC increase).

Then you can run the following command to reproduce the Excel data of Fig.~\ref{figure:overall_compare}. The Fig\_5.csv file, generated in the current directory, can be validated through comparison with the Fig\_5.csv file, which contains the data used in Fig.~\ref{figure:overall_compare}, located in the expected\_results folder.
\begin{lstlisting}[]
  $ python3 Fig5_reproduce.py <output_dir>/compare.csv
\end{lstlisting}
Please note that ``\textless{}output\_dir\textgreater{}'' refers to the address where the ``compare.csv'' file is stored, which is the folder named with a timestamp within the `Compare\_log` directory like ./compare\_log/2023\_11\_08\_10\_56\_52/compare.csv.
\subsection{Evaluation and expected results}
There are two key data results: the optimal architecture discovered by DSE (2, 36, 144GB/s, 32GB/s, 16GB/s, 2MB, 1024) and the improvement ratio of G-ARCH+G-MAP compared to S-ARCH+T-MAP (1.98$\times$ performance improvement, 1.41$\times$ energy efficiency improvement, and 14.3\% MC increase). The optimal architecture and improvement ratio will be displayed in the windows after executing dse.sh and compare.sh scripts respectively, and can be verified against the mentioned results. 
Moreover, the breakdown of energy consumption and delay data for Fig.~\ref{figure:overall_compare} can be verified by comparing the ``Fig\_5.csv'' file generated in GEMINI folder with the ``Fig\_5.csv'' located in the ``expected\_results'' directory. 


\subsection{Methodology}

Submission, reviewing and badging methodology:

\begin{itemize}
  \item \url{https://www.acm.org/publications/policies/artifact-review-badging}
  \item \url{http://cTuning.org/ae/submission-20201122.html}
  \item \url{http://cTuning.org/ae/reviewing-20201122.html}
\end{itemize}


\end{document}